\documentclass[preprint,aps,superscriptaddress,nofootinbib,longbibliography]{revtex4-1}
\usepackage[T1]{fontenc}
\usepackage[latin9]{inputenc}
\usepackage{amsmath}
\usepackage{graphicx}
\usepackage{babel}
\usepackage{url}
\usepackage{hyperref}
\usepackage{slashed}
\usepackage{amsthm}
\usepackage{xcolor}
\newtheorem*{theorem*}{Theorem}

\begin{document}
	\title{New Physics in Neutrino Oscillation: Nonunitarity or Nonorthogonality?}
	
	\newcommand{\affUFABC}{Centro de Ci\^encias Naturais e Humanas\;\;\\
		Universidade Federal do ABC, 09.210-170,
		Santo Andr\'e, SP, Brazil}
	
	\author{Chee Sheng Fong}
	\email{sheng.fong@ufabc.edu.br}
	\affiliation{\affUFABC}

\begin{abstract}
Neutrino oscillation phenomenon is a definite evidence of physics beyond the Standard Model (SM) and high precision measurement of neutrino properties will certainly give us clue about what lies beyond the SM. 
In particular, precise measurements of the mixing matrix elements $U_{\alpha i}$ which relate the neutrino flavor $\alpha$ and mass $i$ eigenstates are crucial since 
new physics at scale beyond experimental reach can lead to a \emph{nonunitary} $U$.
This in turns results in \emph{nonorthogonal} neutrino flavor states. How to calculate the oscillation probability in this scenario is an important theoretical issue that will be treated here. We show that probability constructed using theory of projection probability will ensure that the theory remains \emph{unitary} in time evolution and the probabilities of neutrino of certain flavor being detected as all possible flavor states always sum up to unity. 
This result is crucial for discovery of new physics through neutrino oscillation phenomena.

\end{abstract}

\maketitle
\flushbottom

\section{Introduction}

Neutrino mass is a definite evidence of physics beyond the Standard Model (SM). It cannot be overemphasized the importance of scrutinizing the neutrino sector to as high precision as our current and future technology would allow~\cite{Acero:2022wqg,Arguelles:2022tki}, as this will most certainly lead us towards new physics that not only give rise to neutrino mass but address more fundamental questions like why only the left-handed fields feel the weak force, why is the weak scale so much smaller than the Planck scale, why is electric charge quantized and so on. From the Effective Field Theory (EFT) point of view, even if the new physics scale is beyond our experimental reach, with high enough precision, we will learn important clues about the new physics.

Treating the SM as an EFT, neutrino mass arises, after the electroweak symmetry breaking, from the unique dimension-5 Weinberg operator~\cite{Weinberg:1979sa,Weinberg:1980bf}
\begin{eqnarray}
	{\cal O}_{5} & = & \frac{\lambda_{\alpha\beta}}{\Lambda_{5}}\left(\overline{L_{\alpha}^{c}}\epsilon H\right)\left(L_{\beta}^T\epsilon H\right),\label{eq:O5}
\end{eqnarray} 
where $L_\alpha (\alpha = e, \mu, \tau)$ and $H$ are the $SU(2)_{L}$ lepton and Higgs doublets, respectively, and $\epsilon$ is an $SU(2)_L$ antisymmetric tensor used to contract the two doublets to form a singlet. Here $\lambda$ is a dimensionless symmetric matrix, and $\Lambda_{5}$ is the new physics scale below which the operator ${\cal O}_{5}$ is valid. The existence of this operator is in good agreement with experimental observations in neutrino oscillation phenomena~\cite{Esteban:2020cvm} while the other key prediction is in neutrinoless double-beta decay which might be discovered in current or future experiments~\cite{Agostini:2022zub,Cirigliano:2022oqy}. Next, considering the following dimension-6 operator~\cite{Antusch:2006vwa,Broncano:2002rw,Antusch:2014woa,Fernandez-Martinez:2016lgt}
\begin{eqnarray}
	{\cal O}_{6} & = & \frac{\eta_{\alpha\beta}}{\Lambda_{6}^{2}}\left(\overline{L_{\alpha}}\epsilon H^{*}\right)i\slashed{\partial}\left(L_{\beta}^T\epsilon H\right),\label{eq:O6}
\end{eqnarray}
where $\eta$ is a dimensionless Hermitian matrix and $\Lambda_6$ is the new physics scale (which might or might not be related to $\Lambda_5$) below which the operator is valid. The kinetic term of the SM neutrinos will be modified after the electroweak symmetry breaking. Once neutrino fields are redefined such that kinetic terms are again canonical, the matrix $U$ which relates the neutrino mass and flavor eigenstates is no longer unitary. 
Implicitly, we are assuming that the center of mass energy in an experiment $E\ll\Lambda_{5},\Lambda_{6}$ such that there is no new degrees of freedom beyond those of the SM.
As a remark, not all ultraviolet models which generate ${\cal O}_{5}$ also generate ${\cal O}_{6}$. For instance, type-I and type-III seesaw models generate both ${\cal O}_{5}$ and ${\cal O}_{6}$ while type-II seesaw model only generates ${\cal O}_{5}$ (see for example a review article~\cite{deGouvea:2016qpx}). Here we clearly see the importance of measuring $U$ to a very high precision as it gives us a clue what UV models can give rise to modifications in neutrino sector and the scale in which they reside. Clearly, this cannot be achieved without a precise theoretical formalism of how to deal with possible nonunitary $U$.

Our main result is to show that the net effect of a nonunitary $U$ is that the neutrino flavor states become \emph{nonorthogonal} while the theory remain unitary, as opposed to what have been studied so far in both phenomenological and experimental work involving neutrino oscillation phenomena with nonunitary $U$~\cite{Antusch:2006vwa,Fernandez-Martinez:2007iaa,Xing:2007zj,Goswami:2008mi,Antusch:2009pm,Xing:2011ur,Escrihuela:2015wra,Parke:2015goa,Dutta:2016vcc,Dutta:2016czj,Ge:2016xya,Blennow:2016jkn,Martinez-Soler:2018lcy,Martinez-Soler:2019noy,Dutta:2019hmb,Ellis:2020hus,Wang:2021rsi,Forero:2021azc,Denton:2021mso,Agarwalla:2021owd,Majumdar:2022nby}. With hindsight, this surprising conclusion is perhaps to be expected from the outset since without new degrees of freedom accessible in experiments, i.e. the flavor states are complete, the theory should remains unitary.\footnote{This is in contrast to the case where kinematically accessible light sterile neutrinos which mix with the SM neutrinos are produced in the experiments and lead to apparent unitarity violation due to additional states that are not detected~\cite{Fong:2016yyh,Fong:2017gke,Fong:2022oim,Fong:2023fpt}.}

It is interesting to note that nonorthogonal basis states are commonplace in quantum chemistry~\cite{Mulliken:1955,Roby:1974,Leon:1988,Manning:1990,Artacho:1991,Soriano:2014,Artacho:2017} since it is convenient to express molecular orbitals as linear combinations of atomic orbitals which are in general not orthogonal. 
In this study, nonorthogonal neutrino flavor states are imposed on us due to new physics which results in nonunitary $U$. While the theory is \emph{unitary}, we will continue to use the term \emph{high scale unitarity violation} (in reference to the mixing matrix $U$) to describe this scenario.

\section{Neutrino effective field theory}\label{sec:models}

Here we will assume the center-of-mass energy $E$ involved is below the electroweak symmetry breaking $E<v_{\textrm{EW}}\equiv174$ GeV and also $E < \Lambda_5, \Lambda_6$. Due to the operators \eqref{eq:O5} and \eqref{eq:O6}, the general neutrino Lagrangian allowed by the SM electromagnetic gauge symmetry $U(1)_{\textrm{EM}}$
in the charged lepton mass basis is given by\footnote{The lepton flavors are defined by the corresponding charged lepton masses.}
\begin{eqnarray}
&&{\cal L}_{\nu} = \frac{1}{2}\left(i\overline{\nu_{\alpha}}\slashed{\partial}D_{\alpha\beta}\nu_{\beta}-\overline{\nu_{\alpha}^{c}}m_{\alpha\beta}\nu_{\beta}+{\rm h.c.}\right)\nonumber \\
&&-\left(\frac{g}{2}W_{\mu}^{-}\overline{\ell_{\alpha}}\gamma^{\mu}P_{L}\nu_{\alpha}+\frac{g}{\sqrt{2}\cos\theta_{W}}Z_{\mu}\overline{\nu_{\alpha}}\gamma^{\mu}P_{L}\nu_{\alpha}+{\rm h.c.}\right),
\label{eq:Lag_high_UV}
\end{eqnarray}
where the flavor indices are $\alpha,\beta=e,\mu,\tau$, $D = I + \eta v_{\textrm{EW}}^2/\Lambda_6^2$ with $I$ the $3\times3$ identity matrix, $m = \lambda v_{\textrm{EW}}^2/\Lambda_5$, $g$ is the $SU(2)_{L}$ gauge
coupling, $\theta_{W}$ is the weak angle, $P_{L}=\frac{1}{\sqrt{2}}\left(1-\gamma^{5}\right)$
is the left-handed projector, $\left\{ \ell_{e},\ell_{\mu},\ell_{\tau}\right\} \equiv\left\{ e^{-},\mu^{-},\tau^{-}\right\} $
are the charged leptons and $W^{\mp}$ and $Z$ are the charged and neutral weak bosons, respectively. 

A canonical normalized kinetic term can be obtained by diagonalizing the kinetic term as $D=Y^{\dagger}\hat{D}Y$ where $Y$ is unitary and $\hat{D}$ is real positive and diagonal and then redefining the neutrino fields as $\widetilde{\nu}=\sqrt{\hat{D}}Y\nu$. Next the symmetric mass matrix $\widetilde m \equiv \sqrt{\hat{D}}^{-1}Y^{*}mY^{\dagger}\sqrt{\hat{D}}^{-1}$ can be diagonalized by a unitary matrix $V$ as $\widetilde{m}=V^{*}\hat{m}V^{\dagger}$ where $\hat{m}$
is real and diagonal. Denoting the neutrino fields in the mass basis as $\hat{\nu} \equiv V^{\dagger}\widetilde{\nu}$, we have~\cite{Antusch:2006vwa}
\begin{eqnarray}
{\cal L}_{\nu} & = & \frac{1}{2}\left(i\overline{\hat{\nu}_{i}}\slashed{\partial}\hat{\nu}_{i}-\overline{\hat{\nu}_{i}^{c}}\hat{m}_{ii}\hat{\nu}_{i}+{\rm h.c.}\right)\nonumber \\
 &  & -\left[\frac{g}{2}W_{\mu}\overline{\ell_{\alpha}}\gamma^{\mu}P_{L}U_{\alpha i}\hat{\nu}_{i}+\frac{g}{\sqrt{2}\cos\theta_{W}}Z_{\mu}\overline{\hat{\nu}_{i}}\left(U^{\dagger}U\right)_{ij}\gamma^{\mu}P_{L}\hat{\nu}_{j}+{\rm h.c.}\right],
 \label{eq:EFT_nu_lag}
\end{eqnarray}
where we denote $i,j=1,2,3$ to be the indices in mass basis and we have defined
\begin{eqnarray}
U_{\alpha i} & \equiv & \left(Y^{\dagger}\sqrt{\hat{D}}^{-1}V\right)_{\alpha i},
\end{eqnarray}
which is not unitary in general
\begin{eqnarray}
UU^{\dagger} & = & Y^{\dagger}\hat{D}^{-1}Y,\qquad U^{\dagger}U=V^{\dagger}\hat{D}^{-1}V.
\end{eqnarray}
unless $\hat{D}=I$.

\section{Neutrino oscillation with nonorthogonal flavor basis}\label{sec:theory}

From \eqref{eq:EFT_nu_lag}, the neutrino flavor states $\left|\nu_{\alpha}\right\rangle $ are related to the mass eigenstates $\left|\nu_{i}\right\rangle $ as follows
\begin{eqnarray}
	\left|\nu_{\alpha}\right\rangle  & = & \sum_{i} \overline{U}_{\alpha i}^{*}\left|\nu_{i}\right\rangle, \label{eq:mass_to_flavor}
\end{eqnarray}
where we have defined $\overline{U}_{\alpha i} \equiv U_{\alpha i}/\sqrt{\left(U U^{\dagger}\right)_{\alpha\alpha}}$.
From the orthogonality of mass eigenstates $\left\langle \nu_{j}|\nu_{i}\right\rangle =\delta_{ji}$,
the flavor states \eqref{eq:mass_to_flavor} are properly normalized $\left\langle \nu_{\alpha}|\nu_{\alpha}\right\rangle =1$ though they are in general \emph{nonorthogonal}
\begin{eqnarray}
	\left\langle \nu_{\beta}|\nu_{\alpha}\right\rangle  & = & \big(\overline{U}\,\overline{U}^{\dagger}\big)_{\beta\alpha}
	\equiv {\cal N}_{\beta\alpha},
	\label{eq:nonorthogonality}
\end{eqnarray}
where ${\cal N}$ is a $3\times 3$ matrix with diagonal elements all equal to one.
While the orthogonality of mass basis $\left\{ \left|\nu_{i}\right\rangle \right\} $ implies the usual completeness relation
\begin{eqnarray}
	\sum_{i}\left|\nu_{i}\right\rangle \left\langle \nu_{i}\right| & = & \mathbf{I},\label{eq:completeness_mass}
\end{eqnarray}
with $\mathbf{I}$ the identity operator, the nonorthogonality flavor basis \eqref{eq:nonorthogonality} implies a modified completeness relation~\cite{Fong:2023fpt}\footnote{One can take $g^{\alpha\beta} \equiv ({\cal N}^{-1})_{\alpha\beta}$ as the metric which raises the indices of $\left|_\alpha\right\rangle \equiv \left|\nu_\alpha\right\rangle$ as $\left|^\alpha\right\rangle = g^{\alpha\beta} \left|_\beta\right\rangle$ to form the dual vector. We will avoid using this notation in this work.}
\begin{eqnarray}
	\sum_{\alpha,\beta}  \left|\nu_\alpha\right\rangle \left({\cal N}^{-1} \right)_{\alpha\beta} \left\langle \nu_\beta \right| &=& \mathbf{I}.
	\label{eq:completeness_flavor}
\end{eqnarray}

In quantum mechanics, probability is not an observable and there is no associated Hermitian operator. For orthonormal basis, the probability can be determined by inserting the projection operator $\left|\nu_\beta\right\rangle\left\langle\nu_\beta\right|$ in between $\left\langle \nu_{\alpha}|\nu_{\alpha}\right\rangle$ to obtain the probability of finding $\left|\nu_\beta\right\rangle$ in the original $\left|\nu_\alpha\right\rangle$: $P_{\beta\alpha} = |\left\langle \nu_{\beta}|\nu_{\alpha}\right\rangle|^2$ which is the Born rule.
If the complete set $\left\{ \left|\nu_\alpha\right\rangle \right\}$ are not orthogonal, from eq.~\eqref{eq:completeness_flavor}, the projection operator is given by
\begin{eqnarray}
	P_{\alpha} & \equiv & \sum_{\beta}  \left|\nu_\alpha\right\rangle \left({\cal N}^{-1} \right)_{\alpha\beta} \left\langle \nu_\beta \right|,
	\label{eq:general_projection}
\end{eqnarray}
which satisfies $P_{\alpha}^{2}=P_{\alpha}$ and $\sum_{\alpha}P_{\alpha}=\mathbf{I}$.
Inserting $P_\alpha$ in between $\left\langle \nu_{\alpha}|\nu_{\alpha}\right\rangle$, we obtain
\begin{eqnarray}
	\left\langle \nu_\alpha|P_{\beta}|\nu_\alpha\right\rangle  & = & \left|\left\langle \nu_{\beta}|\nu_{\alpha}\right\rangle\right|^{2}
	+\sum_{\gamma\neq\beta}
	\left\langle \nu_{\alpha}|\nu_{\beta}\right\rangle
	\left({\cal N}^{-1} \right)_{\beta\gamma} 
	\left\langle \nu_{\gamma}|\nu_{\alpha}\right\rangle,
\end{eqnarray}
where the second term is in general complex and cannot be interpreted as a probability. Besides being
real and positive, the probabilities of
finding $\left|\nu_\alpha\right\rangle $ in all possible $\left|\nu_\beta\right\rangle$ should
sum up to \emph{unity} as required since the set $\left\{ \left|\nu_\alpha\right\rangle \right\}$ is \emph{complete}. We will discuss this construction in Section~\ref{subsec:osc_prob}.

\subsection{Evolution of neutrino flavor state}
\label{subsec:evolution}

Let us give brief review of evolution of a flavor state. Starting from an initial state $\left|\nu_{\alpha}\left(0\right)\right\rangle =\left|\nu_{\alpha}\right\rangle $,
the time-evolved state $\left|\nu_{\alpha}\left(t\right)\right\rangle $
is described by the Schrödinger equation
\begin{eqnarray}
i\frac{d}{dt}\left|\nu_{\alpha}\left(t\right)\right\rangle  & = & {\cal H}\left|\nu_{\alpha}\left(t\right)\right\rangle ,\label{eq:Schr_eq}
\end{eqnarray}
where the Hamiltonian is ${\cal H}={\cal H}_{0}+{\cal H}_{I}$ with
${\cal H}_{0}$ the free Hamiltonian ${\cal H}_{0}\left|\nu_{i}\right\rangle = E_{i}\left|\nu_{i}\right\rangle$ and $E_{i}=\sqrt{\vec{p}_{i}^{\;~2}+m_{i}^{2}}$,
and ${\cal H}_{I}$ the interaction Hamiltonian with matrix elements $\left\langle \nu_{\beta}\right|{\cal H}_{I}\left|\nu_{\alpha}\right\rangle  = V_{\beta\alpha}$
where $V_{\beta\alpha}^{*}=V_{\alpha\beta}$ since ${\cal H}_{I}^{\dagger}={\cal H}_{I}$.

Multiplying $\left|\nu_{\beta} \right\rangle$ from the left of eq.~\eqref{eq:Schr_eq} and inserting the completeness relations \eqref{eq:completeness_mass} and \eqref{eq:completeness_flavor}, we arrive at
\begin{eqnarray}
	i\frac{d}{dt}\left\langle \nu_{\beta}|\nu_{\alpha}\left(t\right)\right\rangle 
	& = & \sum_{\eta}\left\{\sum_{i}\overline{U}_{\beta i}E_{i}
	\big(\overline{U}^{-1}\big)_{i\eta}
	+\sum_{\gamma}V_{\beta\gamma}\left({\cal N}^{-1} \right)_{\gamma\eta} \right\}
	\left\langle \nu_{\eta}|\nu_{\alpha}\left(t\right)\right\rangle .
\end{eqnarray}
Assuming relativistic neutrinos $E\gg m_{i}$, we expand $E_{i}\simeq E+\frac{m_{i}^{2}}{2E}$ and trade time for distance $t=x$ and obtain, in matrix notation
\begin{eqnarray}
i\frac{dS\left(x\right)}{dx} & = & \left[\overline{U}\Delta\overline{U}^{-1}
+V\left({\cal N}^{-1} \right) \right]S\left(x\right),\label{eq:evol_flavor_basis}
\end{eqnarray}
where we have defined $S_{\beta\alpha}(x) \equiv \left\langle \nu_{\beta}|\nu_{\alpha}\left(x\right)\right\rangle $ and
\begin{eqnarray}
\Delta & \equiv & \frac{1}{2E}\textrm{diag}\left(m_{1}^{2},m_{2}^{2},...,m_{3+N}^{2}\right)=\textrm{diag}\left(\Delta_{1},\Delta_{2},...,\Delta_{3+N}\right).\label{eq:Delta}
\end{eqnarray}
We have dropped the constant $E$ which, as an overall phase in $S(x)$, is not observable.

From eq. (\ref{eq:evol_flavor_basis}), the Hamiltonian in the flavor basis given by
\begin{eqnarray}
H & \equiv & \overline{U}\Delta\overline{U}^{-1}
+V\left({\cal N}^{-1} \right),\label{eq:H_flavor_basis}
\end{eqnarray}
is not Hermitian $H^{\dagger}\neq H$. Through a similarity transformation, we obtain the Hamiltonian in the \emph{vacuum mass basis}
\begin{eqnarray}
\widetilde{H} & \equiv & \overline{U}^{-1}H\overline{U}=\Delta+\overline{U}^{-1}V\overline{U}^{\dagger,-1}.\label{eq:H_mass_basis}
\end{eqnarray}
Since $\widetilde{H}= \widetilde{H}^\dagger$ is Hermitian, it can be diagonalized by a unitary matrix $X$ and has real eigenvalues. So $H$ has the same eigenvalues as $\widetilde H$ and the time evolution of the system is \emph{unitary}. The apparent non-Hermicity of $H$ is just due to nonunitary transformation matrix $U$. 
Up to the eigenvalues, we can solve for $S$ analytically for an arbitrary matter potential as shown in refs.~\cite{Fong:2022oim,Fong:2023fpt}.

\subsection{Oscillation probability}
\label{subsec:osc_prob}

In refs.~\cite{Leon:1988,Manning:1990}, the theory of projected probabilities on nonorthogonal states are developed and applied to determine the atomic populations in molecules. We will follow their procedure to derive the neutrino oscillation probability. Given an arbitrary state $\left|\psi\right\rangle$, the basic idea is to project it to a chosen $\left|\nu_{\alpha}\right\rangle$ and to the corresponding orthogonal component $\left|\nu_{\alpha}\right\rangle_\perp$.
The orthogonal component is further projected to the (hyper)plane formed by the rest of the basis states and to the orthogonal component to this plane. Then the new orthogonal component is further projected to $\left|\nu_{\alpha}\right\rangle$ and $\left|\nu_{\alpha}\right\rangle_\perp$ and so on. It turns out
that for two and three states system, the probability operators have closed forms and we will refer the reader to the companion paper for details~\cite{Fong:2023fpt}.
After solving the amplitude $S(x)$ as in Section~\ref{subsec:evolution}, the probability of an initial state $\left|\nu_{\alpha}\right\rangle $ being detected as $\left|\nu_{\beta}\right\rangle $ at distance $x$ can be written as
\begin{eqnarray}
	P_{\beta\alpha}\left(x\right) & = & \sum_{\xi,\lambda}
	S_{\alpha\xi}(x) (\hat P_{\beta})_{\xi\lambda} S_{\lambda\alpha}(x),
	\label{eq:prob}
\end{eqnarray}
where $\sum_{\beta}(\hat P_{\beta})_{\xi\lambda} = ({\cal N}^{-1})_{\xi\lambda}$ which together with eq.~\eqref{eq:completeness_flavor} guarantees that $\sum_{\beta} P_{\beta\alpha}(x) = 1$ as required by unitarity.
The appearance of $(\hat P_{\beta})_{\xi\lambda}$ (which only depends on $U$) takes into account possible nonorthogonality of flavor states where for vanishing off-diagonal elements of ${\cal N}$, we have
\begin{eqnarray}
	(\hat P_\beta)_{\xi\lambda} & = & \delta_{\xi\beta}\delta_{\beta\lambda},
	\label{eq:standard_result}
\end{eqnarray}
and we recover the standard result. Utilizing eq.~\eqref{eq:standard_result} for nonorthogonal flavor states will lead to inconsistent result $\sum_{\beta} P_{\beta\alpha}(x) \neq 1$. To prove this, it is sufficient to show the case for $x=0$ in which we obtain
\begin{eqnarray}
	P_{\beta\alpha}\left(0\right) & = &
 |{\cal N}_{\beta\alpha}|^2,
\end{eqnarray}
which gives $\sum_{\beta} P_{\beta\alpha}(0) > 1$ since ${\cal N}_{\alpha\alpha} = 1$.

Next we define ${\cal N}_\alpha$ as a $2\times 2$ submatrix formed from the matrix ${\cal N}$ excluding the row and column involving $\nu_\alpha$ state.
Following the construction of refs.~\cite{Leon:1988,Manning:1990}, we obtain our main result (see the companion paper~\cite{Fong:2023fpt} for details)
\begin{eqnarray}
	(\hat{P}_{\alpha})_{\xi\lambda} & = & \frac{1}{3}\left[(E_{\alpha})_{\xi\lambda}
	+\sum_{\beta\neq\alpha}(F_{\alpha\beta})_{\xi\lambda}\right],
	\label{eq:new_result}
\end{eqnarray}
with
\begin{eqnarray}
	\left(E_{\alpha}\right)_{\xi\lambda} & = & \begin{cases}
		{\displaystyle 1 + \frac{X_{\alpha}^{2}}{1-X_{\alpha}^{2}}}, & \xi=\lambda=\alpha\\
		{\displaystyle \frac{\left|{\cal N}_{\alpha\xi}-{\cal N}_{\alpha\gamma\xi}\right|^{2}}{\left(\det {\cal N}_{\alpha}\right)^{2}\left(1-X_{\alpha}^{2}\right)},\;\gamma\neq\left\{ \alpha,\xi\right\} }, & \xi=\lambda\neq\alpha\\
		{\displaystyle -\frac{1}{2}\frac{{\cal N}_{\xi\lambda}-{\cal N}_{\xi\gamma\lambda}}{\det {\cal N}}},\;\gamma\neq\left\{ \alpha,\xi\right\} , & \xi\neq\lambda\;\textrm{and}\;(\xi=\alpha\;\textrm{or}\;\lambda=\alpha)\\
		{\displaystyle \frac{\left({\cal N}_{\alpha\lambda}-{\cal N}_{\alpha\xi\lambda}\right)\left({\cal N}_{\xi\alpha}-{\cal N}_{\xi\lambda\alpha}\right)}{\left(\det {\cal N}_{\alpha}\right)^{2}\left(1-X_{\alpha}^{2}\right)}}, & \xi\neq\lambda\;\textrm{and}\;\left\{ \xi,\lambda\right\} \neq\alpha
	\end{cases},
\end{eqnarray}
and for $\beta\neq\alpha$ and $\gamma\neq\left\{ \alpha,\beta\right\} $
\begin{eqnarray}
	\left(F_{\alpha\beta}\right)_{\xi\lambda} & = & \begin{cases}
		{\displaystyle 1 + \frac{\left|{\cal N}_{\alpha\gamma}\right|^{4}}{1-\left|{\cal N}_{\alpha\gamma}\right|^{4}}
			+\frac{\langle\left(p_{\alpha}\right)_{\left\{ \alpha,\gamma\right\} }\rangle_{\beta\beta} \left|{\cal N}_{\alpha\beta}-{\cal N}_{\alpha\gamma\beta}\right|^{2}}{\left(\det {\cal N}_{\beta}\right)^{2}\left(1-X_{\beta}^{2}\right)}}, & \xi=\lambda=\alpha\\
		{\displaystyle \begin{split}-\frac{1}{2}\frac{1}{1-\left|{\cal N}_{\alpha\gamma}\right|^{4}}\left({\cal N}_{\xi\lambda}-\frac{1+\left|{\cal N}_{\alpha\gamma}\right|^{2}}{2}{\cal N}_{\xi\gamma\lambda}\right)\\
		-\frac{1}{2}\frac{\langle\left(p_{\alpha}\right)_{\left\{ \alpha,\gamma\right\} }\rangle_{\beta\beta} }{1-X_{\beta}^{2}}
				\frac{{\cal N}_{\xi\lambda}-{\cal N}_{\xi\gamma\lambda}}{\det {\cal N}_{\beta}}\left(1+\left\langle p_{\alpha\gamma}\right\rangle_{\beta\beta} \right)
			\end{split}
		}, & \xi\lambda=\alpha\beta\;\textrm{or}\;\xi\lambda=\beta\alpha\\
		{\displaystyle 
			-\frac{1}{2}\frac{{\cal N}_{\xi\lambda}}{\det {\cal N}_{\beta}}
			+\frac{\langle\left(p_{\alpha}\right)_{\left\{ \alpha,\gamma\right\} }\rangle_{\beta\beta} \left({\cal N}_{\xi\beta}-{\cal N}_{\xi\lambda\beta}\right)\left({\cal N}_{\beta\lambda}-{\cal N}_{\beta\xi\lambda}\right)}{\left(\det {\cal N}_{\beta}\right)^{2}\left(1-X_{\beta}^{2}\right)}}, & \xi\lambda=\alpha\gamma\;\textrm{or}\;\xi\lambda=\gamma\alpha\\
		{\displaystyle \frac{1}{2}\left(1+\frac{1+\left\langle p_{\alpha\gamma}\right\rangle_{\beta\beta}^{2}}{1-X_{\beta}^{2}}\right)
			\langle\left(p_{\alpha}\right)_{\left\{ \alpha,\gamma\right\} }\rangle_{\beta\beta} }, & \xi=\lambda=\beta\\
		{\displaystyle \begin{split}-\frac{1}{2}\frac{1}{1-\left|{\cal N}_{\alpha\gamma}\right|^{4}}\left(\left|{\cal N}_{\alpha\gamma}\right|^{2}{\cal N}_{\xi\lambda}-\frac{1+\left|{\cal N}_{\alpha\gamma}\right|^{2}}{2}{\cal N}_{\xi\alpha\lambda}\right)\\
		-\frac{1}{2}\frac{\langle\left(p_{\alpha}\right)_{\left\{ \alpha,\gamma\right\} }\rangle_{\beta\beta} }{1-X_{\beta}^{2}}
				\frac{{\cal N}_{\xi\lambda}-{\cal N}_{\xi\alpha\lambda}}{\det {\cal N}_{\beta}}\left(1+\left\langle p_{\alpha\gamma}\right\rangle_{\beta\beta} \right)
			\end{split}
			,} & \xi\lambda=\beta\gamma\;\textrm{or}\;\xi\lambda=\gamma\beta\\
		{\displaystyle \frac{\left|{\cal N}_{\alpha\gamma}\right|^{2}}{1-\left|{\cal N}_{\alpha\gamma}\right|^{4}}
			+\frac{\langle\left(p_{\alpha}\right)_{\left\{ \alpha,\gamma\right\} }\rangle_{\beta\beta} 
				\left|{\cal N}_{\gamma\beta}-{\cal N}_{\gamma\alpha\beta}\right|^{2}}{\left(\det {\cal N}_{\beta}\right)^{2}\left(1-X_{\beta}^{2}\right)}}, & \xi=\lambda=\gamma
	\end{cases},
\end{eqnarray}
where we have defined
\begin{eqnarray}
	{\cal N_{\alpha\beta\gamma}} & \equiv &
	{\cal N}_{\alpha\beta}{\cal N}_{\beta\gamma}, \\
	\left\langle p_{\alpha\gamma}\right\rangle_{\beta\beta}  & \equiv & \frac{\left|{\cal N}_{\alpha\beta}\right|^{2}+\left|{\cal N}_{\beta\gamma}\right|^{2}
		-2\textrm{Re}\left({\cal N}_{\beta\alpha}{\cal N}_{\alpha\gamma}{\cal N}_{\gamma\beta}\right)}{1-\left|{\cal N}_{\alpha\gamma}\right|^{2}},\\
	\langle\left(p_{\alpha}\right)_{\left\{ \alpha,\gamma\right\} }\rangle_{\beta\beta}  & \equiv & \frac{\left|{\cal N}_{\alpha\beta}\right|^{2}
		+\left|{\cal N}_{\alpha\gamma}{\cal N}_{\beta\gamma}\right|^{2}
		-(1+\left|{\cal N}_{\alpha\gamma}\right|^{2})
		\textrm{Re}\left({\cal N}_{\beta\alpha}{\cal N}_{\alpha\gamma}{\cal N}_{\gamma\beta}\right)}{1-\left|{\cal N}_{\alpha\gamma}\right|^{4}}.
\end{eqnarray}
The subscript $\{\alpha,\gamma\}$ in the last expression are not indices but refers to the set of basis states of the corresponding operator. For example, with $\{\mu,\tau\}$, we can have $\langle\left(p_{\mu}\right)_{\left\{ \mu,\tau\right\} }\rangle_{\beta\beta}$ or $\langle\left(p_{\tau}\right)_{\left\{ \mu,\tau\right\} }\rangle_{\beta\beta}$.
In the case where off-diagonal elements of ${\cal N}$ are zero, we recover the standard result in eq.~\eqref{eq:standard_result}. This also shows that high scale nonunitarity scenario is only sensitive to off-diagonal elements of $U$ and as shown in ref.~\cite{Fong:2023fpt}, in the limit of vanishing off diagonal elements, the scenario is indistinguishable from the unitarity scenario.

\begin{figure}
	\begin{centering}
			\includegraphics[scale=0.4]{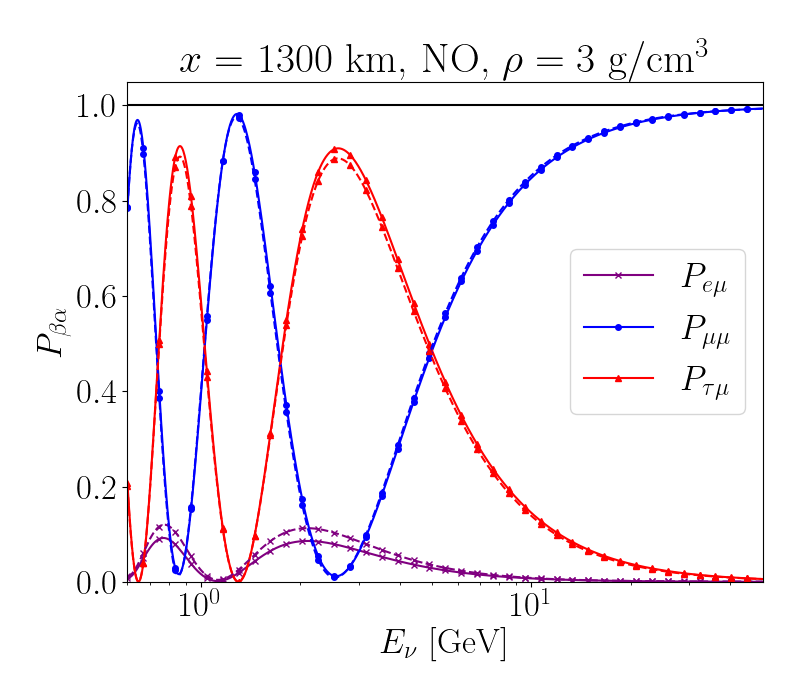}
			\includegraphics[scale=0.4]{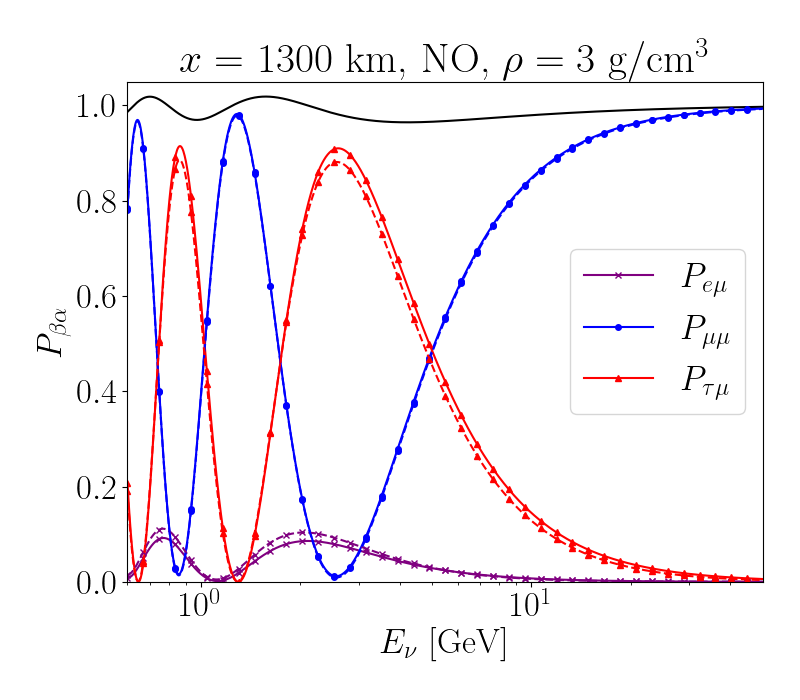}
			\par\end{centering}
	\caption{The probability of $\nu_\mu \to \nu_\beta$ in matter with constant density $\rho = 3\,\textrm{g/cm}^3$ at a distance $x = 1300$ km as a function of neutrino energy $E_\nu$ with $\hat P_\beta$ given by eq.~\eqref{eq:new_result} (left plot) or eq.~\eqref{eq:standard_result} (right plot). The solid curves are for the standard scenario with unitary $U$ while the dashed curves are for nonorthogonal flavor states with $\big(UU^\dagger\big)_{e\mu} = \big(UU^\dagger\big)_{e\tau} = \big(UU^\dagger\big)_{\mu\tau} = 0.03e^{-i\pi/3}$ and $\big(UU^\dagger\big)_{ee} = \big(UU^\dagger\big)_{\mu\mu}= \big(UU^\dagger\big)_{\tau\tau} = 0.96$.
 	The black solid curves at the top denote $\sum_{\beta}P_{\beta\alpha}$ for nonunitary $U$.
    \label{fig:Pxmu_constant_matter}}
\end{figure}

Using the public code \texttt{NuProbe}~\cite{Fong:2022oim,NuProbe}, in Figure~\ref{fig:Pxmu_constant_matter}, we plot the probability of $\nu_\mu \to \nu_\beta$ at a distance of 1300 km with a constant matter density of 3 g/cm$^3$ for $\beta = e,\mu,\tau$ which correspond to purple cross, blue dot and red star curves, respectively. The solid curves are for the standard scenario with unitary $U$ while the dashed curves are for the high scale unitarity violation scenario with nonunitary $U$. The black solid curves on the top denote $\sum_{\beta} P_{\beta\alpha}$ for nonunitary $U$. The standard parameters are set to the global best fit values for Normal mass Ordering (NO) from~\cite{Esteban:2020cvm,NuFIT}. 
The nonorthogonal parameters are set to $\big(UU^\dagger\big)_{e\mu} = \big(UU^\dagger\big)_{e\tau} = \big(UU^\dagger\big)_{\mu\tau} = 0.03e^{-i\pi/3}$ and $\big(UU^\dagger\big)_{ee} = \big(UU^\dagger\big)_{\mu\mu}= \big(UU^\dagger\big)_{\tau\tau} = 0.96$. 
On the left plot, the expression~\eqref{eq:new_result} for $\hat P_\beta$ is used while on the right plot, the expression~\eqref{eq:standard_result} is used. As we can see, with nonunitary $U$, the latter expression gives rise to spurious result in which the total probability can be larger or smaller than 1 (black curve on the right plot). The deviations from the standard scenario are more apparent for appearance channels $\nu_\mu \to \nu_e$ and $\nu_\mu \to \nu_\tau$.

\begin{figure}
	\begin{centering}
		\includegraphics[scale=0.4]{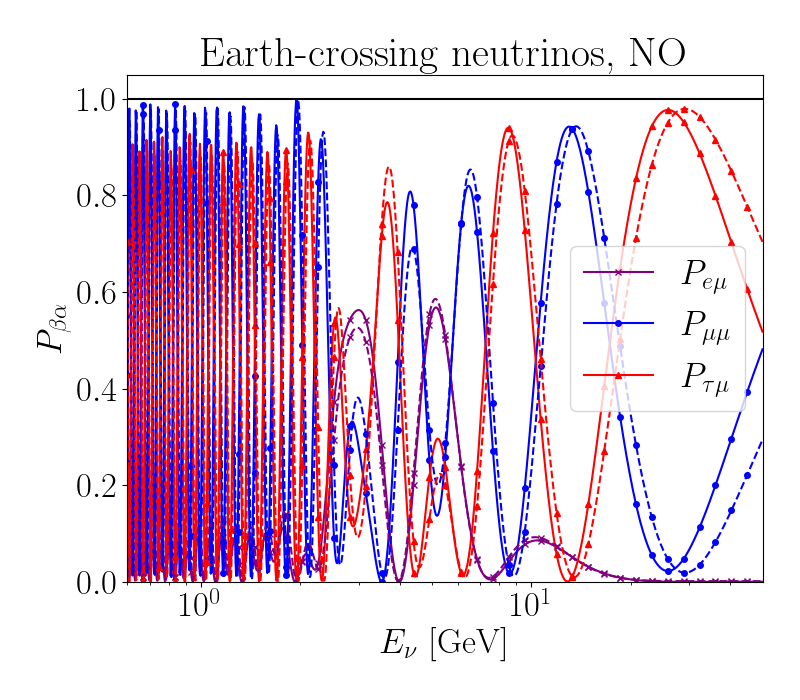}
		\includegraphics[scale=0.4]{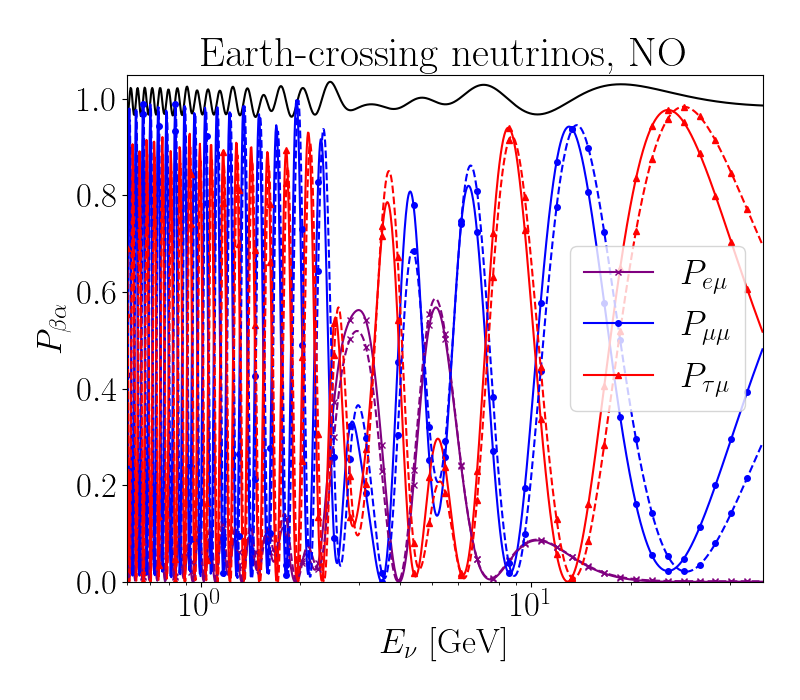}
		\par\end{centering}
	\caption{
		The probability of $\nu_\mu \to \nu_\beta$ for neutrino passing through the Earth core using simplified PREM model as a function of neutrino energy $E_\nu$. The notations are the same as in Figure~\ref{fig:Pxmu_constant_matter}.  \label{fig:Pxmu_earth_crossing}}
\end{figure}

In Figure~\ref{fig:Pxmu_earth_crossing}, we repeat the calculations using the same neutrino parameters but for neutrino crossing through the Earth core with a simplified (Preliminary Reference Earth Model) PREM model~\cite{Dziewonski:1981xy} implemented in \texttt{NuProbe}~\cite{Fong:2022oim,NuProbe}. Even with nontrivial matter potential, unitarity is preserved as can be seen in the black curve on the left plot. We also see that matter effects noticeably enhance the differences between the standard and high scale unitarity violation scenario for all the channels.

\section{Conclusions}\label{sec:conclusions}
To answer the question posted in the title, new physics can result in \emph{nonunitary} lepton mixing matrix $U$ which further implies \emph{nonorthogonal} neutrino flavor states. 
This \emph{high scale unitarity violation} scenario can be distinguished from the standard scenario although the theory itself remains unitary in time evolution and the total probability always sum up to unity.
If new physics scale were beyond the energy reach of our current and foreseeable future experiments, one might have to focus on intensity frontier to carry out precision measurements. In such a scenario, neutrino sector might be the unique place that guarantees clues to new physics. Then precision measurements of neutrino oscillation phenomena rely on precise theoretical treatment and to prepare for discovery of new physics, one should use eq.~\eqref{eq:new_result} in the probability calculation which preserves unitarity.

%%%%%%%%%%%%%%%%%%%%%%%%%%
\section{Acknowledgments}
%%%%%%%%%%%%%%%%%%%%%%%%%%
C.S.F. acknowledges the support by grant 2019/11197-6 and 2022/00404-3 from São Paulo Research Foundation (FAPESP), and grant 301271/2019-4 and 407149/2021-0 from National Council for Scientific and Technological Development (CNPq). 
This work is inspired by a ``dangerous idea'' of Hisakazu Minakata who said ``\textit{A closed theory without unitarity must not make sense. With ignoring inelastic scattering, absorption, and decay, etc, the three neutrino system cannot lose the probability. To where probability loss goes?}''
He would like to thank Celso Nishi for reading and commenting on the manuscript.  He also acknowledges support from the ICTP through the Associates Programme (2023-2028) while this work was being completed.

\bibliography{nuprobe}

\end{document}